\documentclass[intlimits,twoside,a4paper]{article}

\usepackage[cp1251]{inputenc}
\usepackage{multirow}
\usepackage[eqsecnum]{cmpj3}

\issue{2023}{26}{2}{23704}
\doinumber{10.5488/CMP.26.23704}
\title[Electron energy spectrum of the spherical quantum dot]%
{Electron energy spectrum of the spherical GaAs/Al$_x$Ga$_{1-x}$As quantum dot with several impurities on the surface}
\author[R. Ya. Leshko, I. V. Bilynskyi, O. V. Leshko, V. B. Hols'kyi]
{R. Ya. Leshko\orcid{0000-0002-9072-164X}\refaddr{label1}\thanks{Corresponding author: \email{leshkoroman@gmail.com}.},
I. V. Bilynskyi\orcid{0000-0002-4221-9225}\refaddr{label1, label2},
O. V. Leshko\orcid{0000-0001-9646-3189}\refaddr{label1},
V. B. Hols'kyi\orcid{}\refaddr{label1}
}
\addresses{
\addr{label1} Physics Department, Drohobych Ivan Franko State Pedagogical University, 3 Stryiska Street, 82100\\ Drohobych, Ukraine
\addr{label2} Physics Department, Kryvyi Rih State Pedagogical University, 54 Gagarin Avenue, 50086 Kryvyi Rih, Ukraine
}

\date{Received December 4, 2022, in final form January 9, 2023}

\begin{document}

\maketitle

\begin{abstract}
 The model of a spherical quantum dot with several donor impurities on its surface is suggested. The electron energy spectra are studied as a function of the quantum dot radius and the number of impurities.  Several cases of the location of impurities on the quantum dot surface are considered. The plane wave functions method has been applied to calculate the electron energy spectrum. The splitting of electron energy levels is analyzed in the cases of different number of impurities. It is shown that the electron energy splitting depends on both the number of impurities on the surface and on their location. The electron binding energy is defined too.
%
%
\printkeywords
%
\end{abstract}

\section{Introduction}


Recent developments in nanotechnologies have made it possible to produce zero-dimensional semiconductor structures such as quantum dots (QDs). These quantum systems were proposed as solar cells and solar concentrators \cite{d1,d2,d3}, photodetectors \cite{d4,d5,d6}, single QD transistors \cite{d7,d8}, lasers \cite{d9,d10,d11}, and light emitting diodes \cite{d12,d13}. Various methods are used to manufacture these devices based on QD systems. The examples  are molecular beam epitaxy, metal organic chemical vapor deposition, lithography methods, and colloidal methods. Using the above mentioned and other methods does not ensure that QDs are not contaminated by impurities. Obviously, the presence of impurity is undesirable. However, if there is an impurity, it can be located in the QD or on the QD surface. On the other hand, there are many devices, where  doped QDs are used. In this case, impurities are desirable. In both cases, the impurities are mostly  on the QD surface. However, there exists extrinsic and intrinsic doping which provides the presence of impurity not only on the QD surface but also in the QD. For example, a review of colloidal doping methods is presented in \cite{d14}. High quantum yield Cu doped CdSe quantum dots  were studied in \cite{d15}. The authors detected two states of Cu oxidation (+1 and +2) for surface-doped quantum dots. The results revealed that the quantum dots doped with low concentrations were dominated by Cu$^{2+}$ ions, whereas the dots doped with high concentrations were dominated by Cu$^{1+}$ along with a less percentage of Cu$^{2+}$ ions \cite{d15}.

Due to the intensive development of nanofabrication and doping methods, there appeared a lot of theoretical works (some of them had preceded the practical realization) where impurity states were calculated. Special attention is paid to hydrogenic impurities. There are many works regarding the hydrogenic impurities in the QD \cite{d16,d17,d18,d19,d20,d21,d22,d23,d24,d25,d26,d27,d28,d29,d30}. Spherical \cite{d16,d17,d19,d20,d21,d22,d23,d24,d25,d27,d32,d33}, cubic \cite{d31}, parallelepiped-shaped \cite{d26}, lens-shaped \cite{d28}, and ellipsoidal \cite{d18,d29,d30} QDs with impurities were studied. In those works, donor \cite{d16,d17,d18,d19,d20,d21,d22,d23,d25,d26,d28,d29,d30,d31,d33} and acceptor \cite{d24,d27,d32} impurities were considered. Electron and hole spectra, linear and nonlinear optical properties \cite{d18,d19,d22,d28,d30,d33} were analyzed in electric \cite{d25,d26} and magnetic \cite{d21} fields with impurities located in the center and off-center of the QD. Most of those works are devoted to one impurity located in the QD center or outside. A less number of works are connected with two impurities, for example \cite{d32,d33}. However, in all these works the impurities are considered in different QD locations. It is very important to study several impurities located on the QD surface because  many experimental methods make it possibile to dope the surface of the prepared QDs and dope the QD during its growth.

In this work, we consider the cases of several impurities on the QD surface. We consider 1, 2, 4, 6 impurities on the QD surface. Therefore, the aim of this work is
\begin{itemize}
  \item to determine the electron energy in the QD with several impurities on the QD surface;
  \item to establish the influence of the impurity number on the electron spectra;
  \item to calculate the electron binding energy in the QD with surface impurities.
\end{itemize}

\section{Nanoheterosystem model and calculation method}

We consider a semiconductor spherical QD with radius $r_0$, electron effective mass $m_0$ surrounded by a semiconductor matrix with electron effective mass $m_1$. The dielectric constant of the QD and matrix have very close values. Furthermore, we assume that the lattice constants of the QD and matrix have very close values too. That is why we do not regard the influence of polarization and deformation. On the QD surface there are impurity ions. We assume that the surface concentration of impurities is not larger than $\sigma$~=~0.0048~\AA$^{-2}$. For example, if $r_0$~=~20~\AA, $S_{\text{surface}} = \piup r_0^2 \approx 1256$~\AA$^2$, and we get 1256~\AA$^2\cdot$~0.0048~\AA$^{-2} \approx$~6. Therefore, for the QD radius $r_0$~=~20~\AA,~the number of impurities on the QD surface can be 6 or less (if there are 6 impurities, the minimum QD radius should be 20~\AA)

For convenience, we have written the effective mass Hamiltonian of the described system in units of effective Rydberg energy
  and effective Bohr radius:

\begin{align}
\label{Ham}
    {\bf{\hat H}} =  - \nabla \frac{{{m_0}}}{{m\left( r \right)}}\nabla  + U\left( r \right) + V\left( {\vec r,{{\vec D}_1},{{\vec D}_2},...,{{\vec D}_N}} \right),
\end{align}

\noindent where the effective mass

\begin{align}
\label{Mas}
    m\left( r \right) = \left\{ \begin{array}{l}
    {m_0},\,\,\,\,\,r \leqslant {r_0},\\
    {m_1},\,\,\,\,\,\,r > {r_0},
    \end{array} \right.
\end{align}

\begin{align}
\label{Conf}
    U\left( r \right) = \left\{ \begin{array}{l}
    0,\,\,\,\,\,\,\,\,\,\,r \leqslant {r_0},\\
    {U_0},\,\,\,\,\,\,r > {r_0},
    \end{array} \right.
\end{align}
 is a confinement potential, and
\begin{align}
\label{Coul}
    V\left( {\vec r,{{\vec D}_1},{{\vec D}_2},...,{{\vec D}_N}} \right) = \sum\limits_{j = 1}^N {{V_j}\left( {\vec r,{{\vec D}_j}} \right)}  =  - \sum\limits_{j = 1}^N {\frac{2}{{\left| {\vec r - {{\vec D}_j}} \right|}}}
\end{align}

\noindent is Coulomb potential energy of interaction between $j$-th ion (in location $\vec{D}_j$ ) and electron, $N$ is the number of impurities.

To calculate the electron energy spectrum and wave functions, the Schr\"{o}dinger equation should be solved. In the general case of many impurities, this equation cannot be solved exactly. That is why we use the plane wave method, which is described in detail in \cite{d25,d26,d34}. We extend the method to the case of many impurities on the QD surface. The wave function, which is a solution of the Schr\"{o}dinger equation, can be expressed in the form:

\begin{align}
\label{GeneralWave}
    \psi \left( {\vec r} \right) = \sum\limits_{{n_x} =  - {n_{\max }}}^{{n_{\max }}} {\sum\limits_{{n_y} =  - {n_{\max }}}^{{n_{\max }}} {\sum\limits_{{n_z} =  - {n_{\max }}}^{{n_{\max }}} {{C_{{n_x},{n_y},{n_z}}}\psi _{{n_x},{n_y},{n_z}}^{(0)}\left( {x,y,z} \right)} } },
\end{align}

\noindent where $\psi _{{n_x},{n_y},{n_z}}^{(0)}\left( {x,y,z} \right)$ are plane waves which  form a complete closed system of functions in the domain  $L_x \times L_y \times L_z$, $n_{\text{max}} \rightarrow \infty$ (but in numerical calculation,  $n_{\text{max}}$ will be limited),

\begin{align}
\label{Psi0}
    \psi _{{n_x},{n_y},{n_z}}^{(0)}\left( {x,y,z} \right) = \frac{1}{{\sqrt {{L_x} {L_y} {L_z}} }}{{\mathop{\rm e}\nolimits} ^{\ri\left\{ {\left( {{k_x} + {n_x}{K_x}} \right)x + \left( {{k_y} + {n_y}{K_y}} \right)y + \left( {{k_z} + {n_z}{K_z}} \right)z} \right\}}},
\end{align}

\noindent where we use a Cartesian coordinate system and consider the system in the large cube  \noindent ${L_x} = {L_y} = {L_z} \equiv L$, that is why ${K_x} = {K_y} = {K_z} \equiv 2\piup /L$.

After substitution (\ref{GeneralWave}) into the Schr\"{o}dinger equation with Hamiltonian (\ref{Ham}), the linear homogeneous system of equations is obtained:

\begin{align}\label{LinearSystem}
    \sum\limits_{{n_x},{n_y},{n_z} =  - {n_{\max }}}^{{n_{\max }}} {\left[ {{T_{\scriptstyle{{n'}_x},{{n'}_y},{{n'}_z}\hfill\atop
    \scriptstyle{n_x},{n_y},{n_z}\hfill}} + {U_{\scriptstyle{{n'}_x},{{n'}_y},{{n'}_z}\hfill\atop
    \scriptstyle{n_x},{n_y},{n_z}\hfill}} + {V_{\scriptstyle{{n'}_x},{{n'}_y},{{n'}_z}\hfill\atop
    \scriptstyle{n_x},{n_y},{n_z}\hfill}} - E{\delta _{\scriptstyle{{n'}_x},{{n'}_y},{{n'}_z}\hfill\atop
    \scriptstyle{n_x},{n_y},{n_z}\hfill}}} \right]{C_{{n_x},{n_y},{n_z}}}}  = 0,
\end{align}

\noindent where

\begin{align}
     T_{\scriptstyle{{n'}_x},{{n'}_y},{{n'}_z}\hfill\atop
    \scriptstyle{n_x},{n_y},{n_z}\hfill} &=  \left\{ \left( {{k_x} + {{n'}_x}{K_x}} \right)\left( {{k_x} + {n_x}{K_x}} \right) + \left( {{k_y} + {{n'}_y}{K_y}} \right)\left( {{k_y} + {n_y}{K_y}} \right)   \right.
    \nonumber\\
    &+\left. \left( {{k_z} + {{n'}_z}{K_z}} \right)\left( {{k_z} + {n_z}{K_z}} \right)\right\} \times \left\{ {\frac{{{m_0}}}{{{m_1}}}{{{\delta _{\scriptstyle{{n'}_x},{{n'}_y},{{n'}_z}\hfill\atop
    \scriptstyle{n_x},{n_y},{n_z}\hfill}}}} + \frac{{{m_0}}}{{{m_{0,1}}}}S} \right\},
\label{T}
\end{align}

\noindent is the matrix element of the kinetic energy, $m_{0,1}=m_0 m_1 / (m_1-m_0),$
\begin{align}\label{S}
      S = \left\{ \begin{array}{l}
    4\piup r_0^3/(3 L^3)  ,\,\,\,\,\,\,\,\,\,\,\,\,\,\,\,\,\,\,\,\,\,\,\,\,\,\,\,\,\,\,\,\,\,\,\,\,\,\,\,\,\,\,\,\,\,\,\,\,\,\,\,\,\,\,\,\,\,\,\,\,\,\,\,\,\,\,\,\,\,\,{n_x} = {{n'}_x}\,\,\,\,\text{and}\,\,\,\,{n_y} = {{n'}_y}\,\,\,\,\text{and}\,\,\,\,{n_z} = {{n'}_z},\\
    4\piup /(L^3\lambda^3) \cdot \left( {\sin \left( {\lambda {r_0}} \right) - \lambda {r_0}\cos \left( {\lambda {r_0}} \right)} \right),\,\,\,\,{n_x} \ne {{n'}_x}\,\,\,\,\text{or}\,\,\,\,{n_y} \ne {{n'}_y}\,\,\,\,\text{or}\,\,\,\,{n_z} \ne {{n'}_z},
    \end{array} \right.
\end{align}

\begin{align}\label{U}
  {{{U_{\scriptstyle{{n'}_x},{{n'}_y},{{n'}_z}\hfill\atop
    \scriptstyle{n_x},{n_y},{n_z}\hfill}}}} & = {U_0}{{{\delta_{\scriptstyle{{n'}_x},{{n'}_y},{{n'}_z}\hfill\atop
    \scriptstyle{n_x},{n_y},{n_z}\hfill}}}} - {U_0}S,
\end{align}

\noindent is the matrix element of the confinement potential,

\begin{align}\label{V}
  {{V_{\scriptstyle{{n'}_x},{{n'}_y},{{n'}_z}\hfill\atop
    \scriptstyle{n_x},{n_y},{n_z}\hfill}}} = \sum\limits_{j = 1}^N {\left\langle {{{n'}_x},{{n'}_y},{{n'}_z}\big|{V_j}\left( {\vec r,{{\vec D}_j}} \right)|{n_x},{n_y},{n_z}} \right\rangle },
\end{align}

\noindent is the matrix element of the Coulomb potential, where

\begin{align}\label{V-mat}
 & \left\langle {{{n'}_x},{{n'}_y},{{n'}_z}\big|{V_j}\left( {\vec r,{{\vec D}_j}} \right)|{n_x},{n_y},{n_z}} \right\rangle \nonumber\\
  &= \frac{3}{{R_0^3}}{\re^{i\vec \lambda {{\vec D}_j}}}\left\{ \begin{array}{l}
    R_0^2,\,\,\,\,\,\,\,\,\,\,\,\,\,\,\,\,\,\,\,\,\,\,\quad{n_x} = {{n'}_x}\,\,\text{and}\,\,{n_y} = {{n'}_y}\,\,\text{and}\,\,\,{n_z} = {{n'}_z},\\
    \frac{{2 - 2\cos \left( {{R_0}\lambda } \right)}}{{{\lambda ^2}}},\,\,\quad{n_x} \ne {{n'}_x}\,\,\,\text{or}\,\,\,{n_y} \ne {{n'}_y}\,\,\,\text{or}\,\,\,{n_z} \ne {{n'}_z},
    \end{array} \right.
\end{align}

\begin{align}
    {R_0} = L{\left( {\frac{3}{{4\piup }}} \right)^{1/3}},  \,\,\    \vec \lambda  = \frac{{2\piup }}{L}\left[ {\left( {{n_x} - {{n'}_x}} \right){{\vec e}_x} + \left( {{n_y} - {{n'}_y}} \right){{\vec e}_y} + \left( {{n_z} - {{n'}_z}} \right){{\vec e}_z}} \right],
  \nonumber
\end{align}

\noindent where ${\vec e_x},\,\,\,{\vec e_y},\,\,{\vec e_z}$ are unit vectors. In \cite{d25,d26,d34} it was shown that convergence of the results was obtained considering $n_{\text{max}}=7$ and $L=2.5$ $a_b^*+2r_0$. Moreover, in \cite{d34} it was substantiated that with those parameters, the results do not depend on the wave vector $(k_x, k_y, k_z)$ in the range $[0…2\piup/L]$. Our results have a convergence too, even with many impurities for the mentioned parameters.

\section{Analysis of the obtained results}

\begin{figure}[htb]
\centerline{\includegraphics[width=0.9\textwidth]{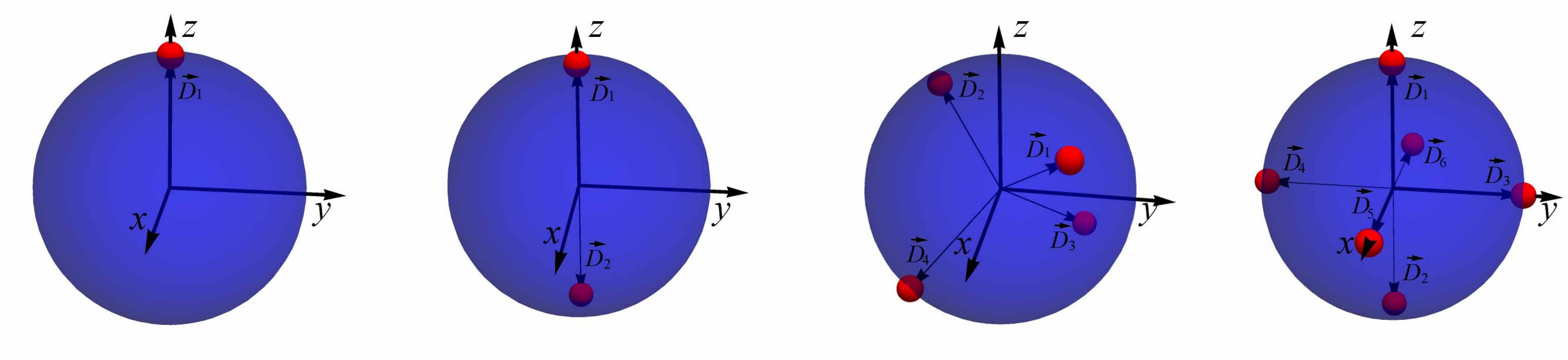}}
\caption{(Colour online) Impurities on the QD surface.} \label{fig1}
\end{figure}

The calculations were performed using the physical parameters of GaAs/Al$_x$Ga$_{1-x}$As semiconductor heterostructure, $x$~=~0.4, $m_0 = 0.067 m_e$, $m_1 = 0.1 m_e$, $U_0$~=~297~meV, $\varepsilon = 13.2$, where $m_e$ is the mass of a free electron in vacuum. We consider four cases of the  location of impurities on the QD surface (figure~\ref{fig1}):

\begin{itemize}
  \item[A)] one impurity $\vec{D_1} = (0,0,r_0)$;
  \item[B)] two impurities $\vec{D_1} = (0,0,r_0)$, $\vec{D_2} = (0,0,-r_0)$;
  \item[C)] four impurities $\vec{D_1} = 1/\sqrt{3} (r_0,r_0,r_0)$, $\vec{D_2} = 1/\sqrt{3} (-r_0,-r_0,r_0)$, $\vec{D_3} = 1/\sqrt{3} (-r_0,r_0,-r_0)$, $\vec{D_4} = 1/\sqrt{3} (r_0,-r_0,-r_0)$;
  \item[D)] six impurities $\vec{D_1} = (0,0,r_0)$, $\vec{D_2} = (0,0,-r_0)$, $\vec{D_3} = (0,r_0,0)$, $\vec{D_4} = (0,-r_0,0)$, $\vec{D_5} = (r_0,0,0)$, $\vec{D_6} = (-r_0,0,0)$.
\end{itemize}

\begin{figure}[htb]
\centerline{\includegraphics[width=1\textwidth]{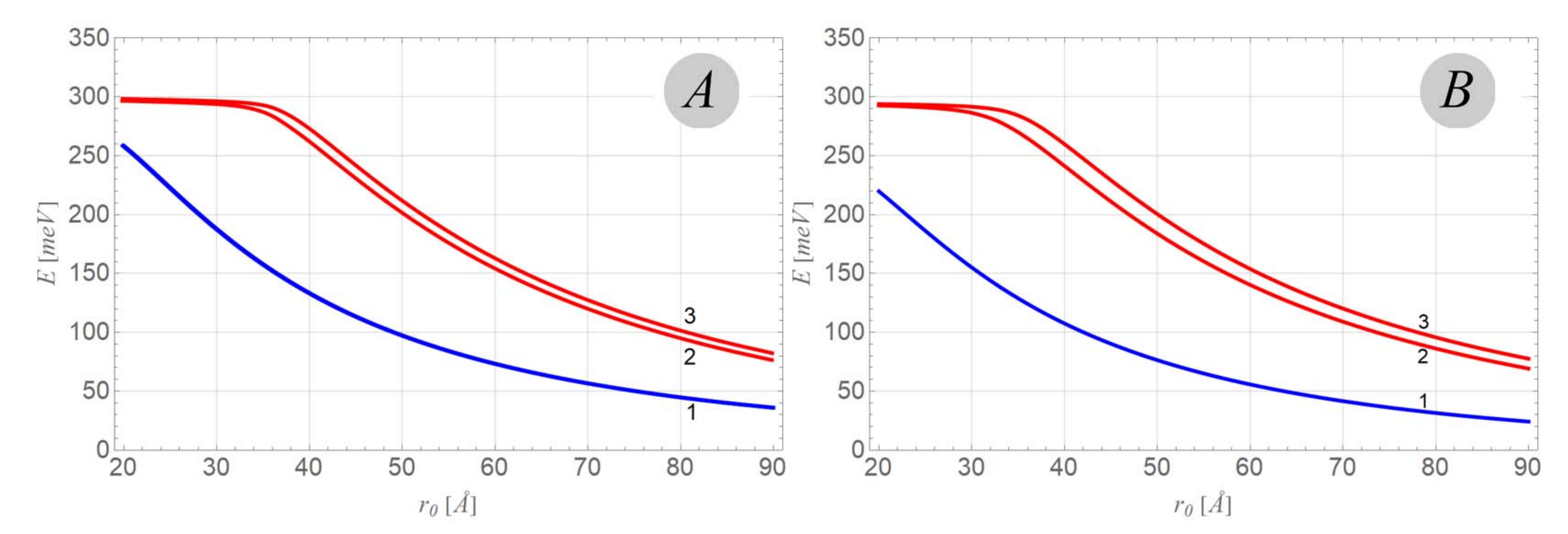}}
\caption{Electron energy spectrum ($s$- and $p$-states) of the QD with one impurity ion (A) and two diametral located ions (B) on the QD surface.}
\label{fig2}
\end{figure}

We perform the calculations for the aforementioned cases in order to define the dependence of several electron energy levels on the QD radius. Moreover, we followed the condition (surface concentration of impurities is not larger than $\sigma$~=~0.0048~\AA$^{-2}$). If there is one impurity ion or two ions on the QD surface, the electron energy spectrum is like the one presented in figure~\ref{fig2}.

\begin{figure}[!t]
\centerline{\includegraphics[width=1\textwidth]{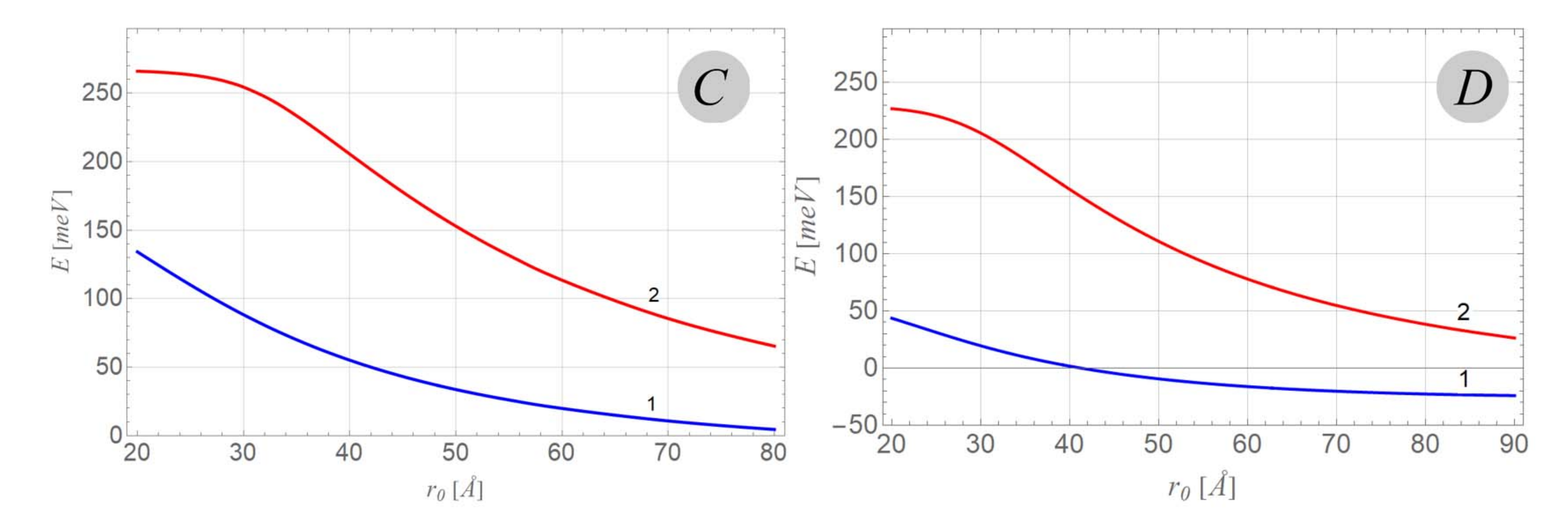}}
\caption{(Colour online) Electron energy spectrum ($s$- and $p$-states) of the QD with 4 impurities ions (C) and 6 ions (D) on the QD surface.}
\label{fig3}
\end{figure}

\begin{table}[!t]
\caption{Electron energy levels in the QD with 4 and 6 impurities on the surface. QD radius is $r_0=70$~\AA.}
\vspace{2mm}

\label{tbl}
\centering
\begin{tabular}{|cc|cc|}
\hline
\multicolumn{2}{|c|}{4 ions of impurity}                                                                                                                                                                       & \multicolumn{2}{c|}{6 ions of impurity}                                                                                                                                                                       \\ \hline
\multicolumn{1}{|c|}{\textit{E}, meV}                                                                & level                                                                                                   & \multicolumn{1}{l|}{\textit{E}, meV}                                                                & level                                                                                                   \\ \hline
\multicolumn{1}{|c|}{10.7}                                                                           & s ($l$ = 0, $m_l$ = 0)                                                                                       & \multicolumn{1}{l|}{$-20.3$}                                                                          & s ($l$ = 0, $m_l$ = 0)                                                                                       \\ \hline
\multicolumn{1}{|c|}{\multirow{3}{*}{\begin{tabular}[c]{@{}c@{}}85.4\\ 85.4\\ 85.4\end{tabular}}}    & \multirow{3}{*}{\begin{tabular}[c]{@{}l@{}}p ($l$ = 1, $m_l = -1$)\\ p ($l$ = 1, $m_l$ = 0)\\ p ($l$ = 1, $m_l$ = 1)\end{tabular}} & \multicolumn{1}{c|}{\multirow{3}{*}{\begin{tabular}[c]{@{}c@{}}54.8\\ 54.8\\ 54.8\end{tabular}}}    & \multirow{3}{*}{\begin{tabular}[c]{@{}l@{}}p ($l$ = 1, $m_l = -1$)\\ p ($l$ = 1, $m_l$ = 0)\\ p ($l$ = 1, $m_l$ = 1)\end{tabular}} \\
\multicolumn{1}{|l|}{}                                                                               &                                                                                                         & \multicolumn{1}{l|}{}                                                                               &                                                                                                         \\
\multicolumn{1}{|l|}{}                                                                               &                                                                                                         & \multicolumn{1}{l|}{}                                                                               &                                                                                                         \\ \hline
\multicolumn{1}{|c|}{\multirow{3}{*}{\begin{tabular}[c]{@{}l@{}}173.9\\ 173.9\\ 173.9\end{tabular}}} & \multirow{3}{*}{\begin{tabular}[c]{@{}l@{}}d ($l$ = 2, $m_l = -1$)\\ d ($l$ = 2, $m_l$ = 0)\\ d ($l$ = 2, $m_l$ = 1)\end{tabular}} & \multicolumn{1}{c|}{\multirow{2}{*}{\begin{tabular}[c]{@{}l@{}}138.5\\ 138.5\end{tabular}}}         & \multirow{2}{*}{\begin{tabular}[c]{@{}l@{}}d ($l$ = 2, $m_l = -2$)\\ d ($l$ = 2, $m_l$ = 2)\end{tabular}}                 \\
\multicolumn{1}{|l|}{}                                                                               &                                                                                                         & \multicolumn{1}{l|}{}                                                                               &                                                                                                         \\ \cline{3-4}
\multicolumn{1}{|l|}{}                                                                               &                                                                                                         & \multicolumn{1}{c|}{\multirow{3}{*}{\begin{tabular}[c]{@{}l@{}}150.7\\ 150.7\\ 150.7\end{tabular}}} & \multirow{3}{*}{\begin{tabular}[c]{@{}l@{}}d ($l$ = 2, $m_l = -1$)\\ d ($l$ = 2, $m_l$ = 0)\\ d ($l$ = 2, $m_l = 1$)\end{tabular}} \\ \cline{1-2}
\multicolumn{1}{|c|}{\multirow{2}{*}{\begin{tabular}[c]{@{}l@{}}180.3\\ 180.3\end{tabular}}}         & \multirow{2}{*}{\begin{tabular}[c]{@{}l@{}}d ($l$ = 2, $m_l = -2$)\\ d ($l$ = 2, $m_l$ = 2)\end{tabular}}                 & \multicolumn{1}{l|}{}                                                                               &                                                                                                         \\
\multicolumn{1}{|l|}{}                                                                               &                                                                                                         & \multicolumn{1}{l|}{}                                                                               &                                                                                                         \\ \hline
\multicolumn{1}{|c|}{216.6}                                                                          & s ($l=0$, $m_l$=0)                                                                                           & \multicolumn{1}{c|}{187.4}                                                                          & s ($l=0$, $m_l$=0)                                                                                      \\ \hline
\end{tabular}
\end{table}

From figure~\ref{fig2} one can see that the electron energy levels decrease when the QD radius increases. In the case of one impurity, we can also use the method of the Schr\"{o}dinger equation solution presented in \cite{d22}. In the case of two diametrally located ions on the QD surface, we also used the method of Schr\"{o}dinger equation solution presented in \cite{d33}. All the methods yield the same results for one and two ions of impurity, respectively. The divergence is no more than 5\%. This comparison of the obtained results with the results obtained by other methods \cite{d22,d33} makes it possible to label the levels in figure~\ref{fig2}: 1 -- first $s$-state; 2,3 -- first $p$-states (magnetic quantum number $m_l = 0, \pm 1$). The splitting of $p$-states can be explained by the violation of spherical symmetry (when impurities are not in the QD center). This splitting is present for all QD radii. For small QD radii, the splitting is very small but still present. Like in our previous works \cite{d22,d33}, the presence of two diametrally located ions on the QD surface causes a smaller electron energy than in the case of one ion on the QD surface. These two cases A) and B) demonstrate that the plane wave method can be successfully used for C) and D) cases of the impurities located on the QD surface.

\begin{figure}[htb]
\centerline{\includegraphics[width=0.6\textwidth]{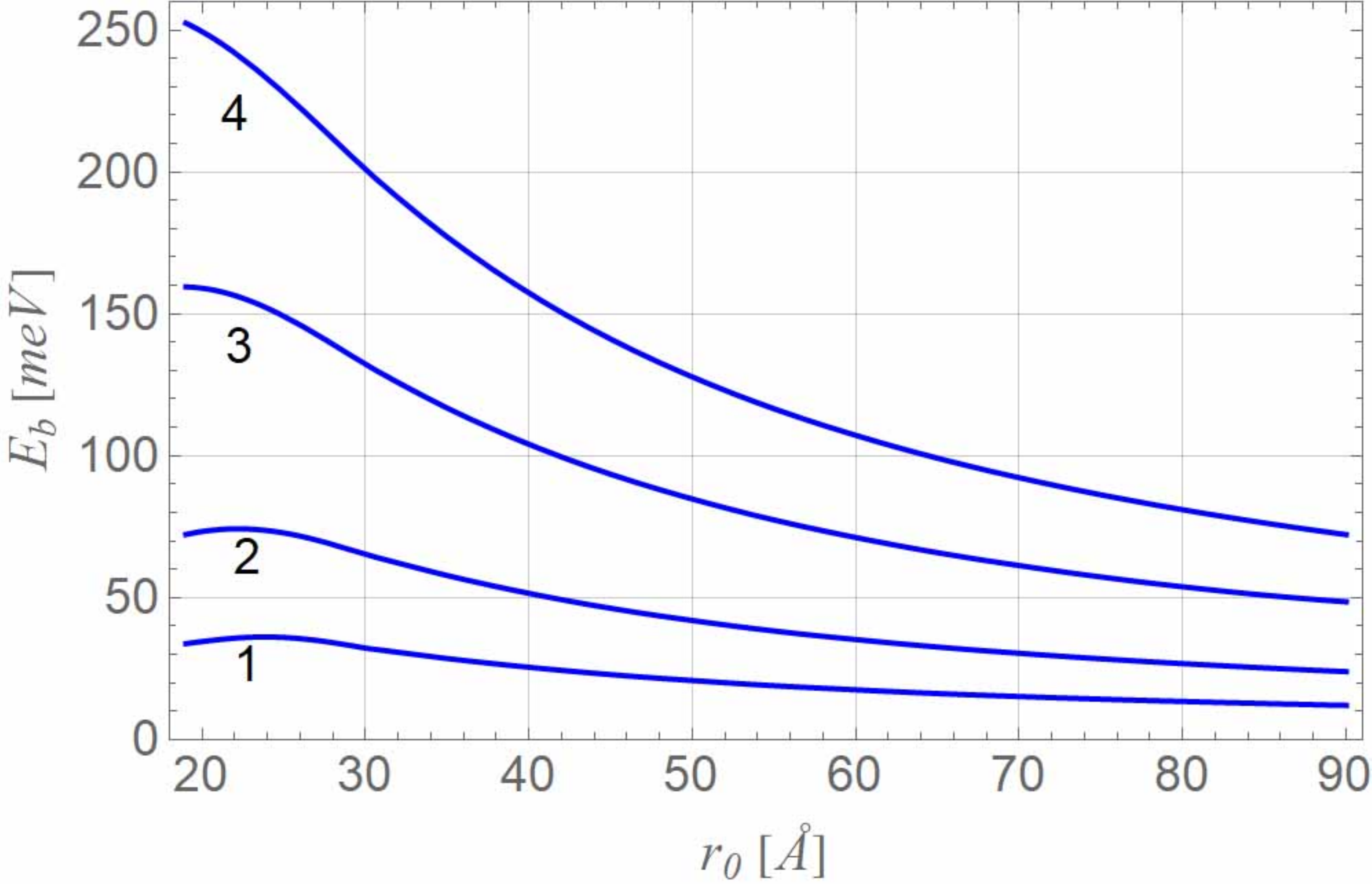}}
\caption{(Colour online) Electron binding energy with impurities. 1--one, 2--two, 3--four, 4--six impurities on the QD surface.}
\label{fig4}
\end{figure}

The results of calculations are presented in figure~\ref{fig3} (C and D cases) and in table~\ref{tbl}. From figure~\ref{fig3} we can see the first $s$-level (curves 1), and the first $p$-level (curves 2). When there are 4 or 6 impurities on the QD surface, we do not observe a splitting of $p$-levels in the figure~\ref{fig3} (see also table~\ref{tbl}). In the cases of one (case A) or 2 (case B) diametrally located impurities, $p$-levels split. Therefore, we can make a conclusion that in the C) and D) cases for $p$-levels, the symmetry of the location of impurities does not cause a $p$-level splitting.

Let us see what happens with $d$-like levels. In the A) and B) cases, we also got a splitting of $d$-like levels~\cite{d24,d33} by $|m_l|=0, 1,2$ (we got 3 levels). Those results are explained by the violation of spherical symmetry and by the existance of cylindrical symmetry. For cylindrical symmetry, a ``good quantum number'' is $|m_l|$. Therefore, the number of the level splitting depends on the possible number of values $|m_l|$. In the C) case, there is tetrahedral symmetry, and the crystal field theory and symmetrical analysis \cite{d35} ``say'' that $d$-levels should be split into two levels: a) twofold degenerate level; b) threefold degenerate level. In the D) case, there is octahedral symmetry and according to the crystal field theory and symmetrical analysis \cite{d35}, $d$-level should be split into two levels: a) threefold degenerate level; b) twofold degenerate level. One can see that in the C) and D) cases the structure of $d$-level splitting is the same, but the order of the splitting levels  is different. Moreover, in both octahedral symmetry and tetrahedral symmetry field, the $p$-level does not split (which can be seen in figure~\ref{fig3}). The results are presented in  table~\ref{tbl}.

In this paper, we also calculate the ground state electron binding  energies for the case of 1, 2, 4, 6 impurities on the QD surface. The calculation results are presented in figure~\ref{fig4}. One can see that the increase of impurities on the QD surface causes an increase of the binding energy due to the presence of additional potential energy [$N$ increase in (\ref{V})].

\section{Conclusion}

In this work we have used the plane wave functions method for the calculation of the electron energy levels in the spherical QD with several impurities on its surface. The obtained results show that for 1 and 2 diametrally located impurity ions, there is a cylindrical symmetry, and we get the energy levels splitting by $|m_l|$. In the case of 4 impurities located in the vertices of a regular tetrahedron on the QD surface and in the case of 6 impurities located on the vertices of regular octahedron, $s$- and $p$-levels do not split. But $d$-levels split into: one threefold degenerated level and one twofold degenerate level (for 4 impurities); one twofold degenerate level and one threefold degenerated level (for 6 impurities). The number of split $d$-levels is the same, but the order is  opposite. Moreover, we have shown that for a larger number of surface impurities, the binding energy increases.

The proposed methodology can be used and extended to non-spherical closed and open \cite{d36} QDs with donor and acceptor impurities on their surface. These calculations will be performed in our next works.


\newpage
\ukrainianpart

\title{Енергетичний спектр електрона у сферичній квантовій точці GaAs/Al$_x$Ga$_{1-x}$As з декількома домішками на поверхні}
\author{Р.~Я.~Лешко\refaddr{label1}, І.~В.~Білинський\refaddr{label1,label2}, О.~В.~Лешко\refaddr{label1}, В.~Б.~Гольський\refaddr{label1}}
\addresses{
\addr{label1} Кафедра фізики, Дрогобицький державний педагогічний університет імені Івана Франка, вул.~Стрийська~3, 82100 Дрогобич, Україна
\addr{label2} Кафедра фізики, Криворізький державний педагогічний університет, проспект Гагаріна, 50086 Кривий Ріг, Україна
}
\makeukrtitle

\begin{abstract}
\tolerance=3000%
Запропоновано модель сферичної квантової точки з декількома донорними домішками на її поверхні. Визначено енергетичний спектр електрона як функцію радіуса квантової точки. Розглянуто декілька випадків розташування домішок на поверхні. Застосовано метод розкладу за плоскими хвилями для обчислення спектру електрона. Проаналізовано розщеплення електронних рівнів для різних випадків розташування домішок. Показано, що енергія розщеплення залежить як від кількості домішок на поверхні, так і від їхнього розташування. Також визначено енергію звязку електрона з домішками.
\keywords енергетичний спектр, поверхнева домішка, метод плоских хвиль

\end{abstract}

\end{document}